# scientific reports

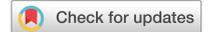

# OPEN  Self-consistent solution for the magnetic exchange interaction mediated by a superconductor

Atousa Ghanbari, Vetle K. Risinggård & Jacob Linder✉

We theoretically determine the magnetic exchange interaction between two ferromagnets coupled by a superconductor using a tight-binding lattice model. The main purpose of this study is to determine how the self-consistently determined superconducting state influences the exchange interaction and the preferred ground-state of the system, including the role of impurity scattering. We find that the superconducting state eliminates RKKY-like oscillations for a sufficiently large superconducting gap, making the anti-parallel orientation the ground state of the system. Interestingly, the superconducting gap is larger in the parallel configuration than in the anti-parallel configuration, giving a larger superconducting condensation energy, even when the preferred ground state is anti-parallel. We also show that increasing the impurity concentration in the superconductor causes the exchange interaction to decrease, likely due to an increasing localization of the mediating quasiparticles in the superconductor.

The Ruderman–Kittel–Kasuya–Yosida (RKKY) is an indirect exchange interaction between localized spins mediated by itinerant electrons in metals[1]. This interaction played an important role in the discovery of giant magnetoresistance (GMR)[2,3] and has been studied in numerous materials[4–9].

The combination of magnetic and superconducting materials has been widely studied due to interesting features which cannot be observed in separate materials[10–19]. Recently, the influence of superconductivity on the magnetic state was experimentally studied in a superconducting spin valve (SSV), GdN-Nb-GdN[20]. On the basis of the de Gennes model[21], it was shown that the superconductor promoted an anti-parallel configuration as the ground-state configuration. In the de Gennes model, a superconductor in an anti-parallel SSV has a higher critical temperature $T_c$ than in the parallel orientation, leading to a larger superconducting gap in the anti-parallel configuration.

The interaction between localized magnetic moments through dirty $s$-wave superconductors[22–24] has previously been found to contain two contributions. One contribution is from the usual RKKY interaction and a second contribution from a longer ranged interaction, decaying exponentially over the superconducting coherence length $\xi$ and with a weaker power-law suppression, which favors an antiferromagnetic alignment. Later, the interaction through a $d$-wave superconductor with an anisotropic order parameter was studied[25]. It was shown on the basis of analytical approximations that this interaction, similarly to the $s$-wave case, contains one oscillatory term and one term favoring an anti-parallel configuration. The oscillations occur when the length of the superconductor ($L_S$) is smaller than the coherence length ($\xi$) while the term favoring an anti–ferromagnetic configuration of the system occurs when $L_S > \xi$. The latter term was found to be proportional to the superconducting gap. Very recently, it was experimentally shown that in a $d$-wave SSV, the anti-parallel ground-state was favored for some specific lengths of the superconducting system and that nodal quasiparticles likely played a central role in mediating the magnetic coupling[26].

In this work, we address numerically and, importantly, self-consistently the effect of conventional $s$-wave singlet superconductors on the indirect exchange coupling ($J$) between two ferromagnetic contacts in a F–S–F structure. The free energy of the system, which is the main quantity of interest in this work, is manifested in a clear experimental observable: namely, the ground-state magnetic configuration of the system. Previous works considering superconducting spin-valves have also considered the superconducting transition temperature as a quantity of interest with respect to possible cryogenic applications. However, a shift in the preferred magnetic

Department of Physics, Center for Quantum Spintronics, Norwegian University of Science and Technology, 7491 Trondheim, Norway. ✉email: jacob.linder@ntnu.no

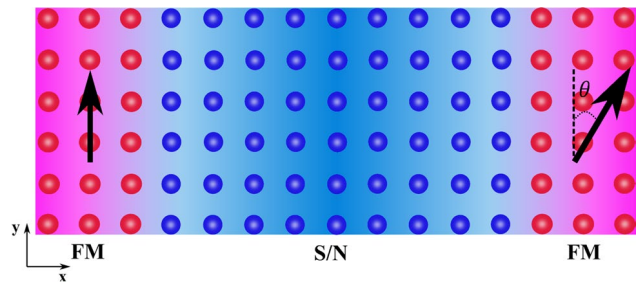

**Figure 1.** Schematic illustration of a superconducting spin valve (SSV) with magnetic moment of the two ferromagnets aligned at relative angle $\theta$. Two cases will be consider here, one with $\theta = 0°$ which is called P orientation and the other one is AP orientation with $\theta = 180°$.

orientation (parallell (P : $\theta = 0°$) or antiparallell (AP : $\theta = 180°$)) of the spin-valve will also be a relevant quantity in this regard. Therefore, determining how the interaction between the ferromagnets depends on the superconducting layer is a task which is both of fundamental and possible practical interest. In contrast to Refs.[22–25], we compute the order parameter self-consistently and account for both the superconducting proximity effect in the ferromagnets and the magnetic proximity effect in the superconducting region. The self-consistent computation of the superconducting order parameter includes not only the effect of the magnetic configuration on the local spin density of the quasiparticles mediating the RKKY interaction, but also the effect of the magnetic configuration on the magnitude of the superconducting gap. This is an important result because as we show in the results section, the difference in gap-magnitude between the parallel and anti-parallel configuration does affect the RKKY-coupling between the ferromagnets. Due to the proximity effect between the superconductors and the ferromagnet, the superconducting gap can be strongly affected by the magnetic configuration, and thus requires a self-consistent calculation, unlike Refs.[22–24] that considered isolated magnetic impurities. In a singlet superconductor, electrons with zero total spin and opposite momentum constitute the Cooper pairs: $(\mathbf{k} \uparrow, -\mathbf{k} \downarrow)$. These Cooper pairs can penetrate into a weak ferromagnet (FM) which has been brought in contact with the superconductor[27] in an oscillatory fashion. Bringing another ferromagnetic layer in contact with this bilayer makes the SSV.

We first briefly reproduce the well-known RKKY-like oscillations of an F–N–F system to contrast these results with what happens in the superconducting state. We consider a finite size system in two dimensions, meaning that we do not assume periodic boundary conditions in any direction. Then, by substituting the central part with a singlet superconductor which leads to a F–S–F structure (Fig. 1), we demonstrate that two types of behaviour take place. For thin superconductors, $J$ oscillates around zero whereas for thick ones the coupling takes values $J > 0$, favoring the AP configuration ($\theta = 180°$), and reduces monotonically as the length is further increased. When the central part is a superconductor with small gap connected to two weakly polarized ferromagnets, we only find RKKY-like oscillations mediated by quasiparticles in the superconductor. In contrast, when the superconducting gap is large or if the exchange field in the ferromagnet is strong, $J > 0$ and the interaction displays either a pure monotonic decay or with superimposed oscillations.

Afterwards, we consider the effect of impurities on $J$ in the F–S–F spin valve. When considering the impurity average $\langle J \rangle_{\text{imp}}$ for a large number of realizations with random impurity configurations, we find that increasing the impurity concentration in the superconductor causes the exchange interaction to decrease. This is likely due to an increasing localization of the mediating quasiparticles in the superconductor[28,29].

## Theory

The indirect exchange interaction between the ferromagnets in F–N–F or F–S–F structures is defined by $J = F^{\uparrow\uparrow} - F^{\uparrow\downarrow}$. Here, $F^{\uparrow\uparrow}$ is the free energy when the ferromagnetic contacts have a parallel (P) orientation ($\theta = 0°$) and $F^{\uparrow\downarrow}$ is the free energy when they have an anti-parallel (AP) orientation ($\theta = 180°$). In this work, we only consider P and AP configurations as the possible ground states. This assumption is possible as we will consider strong anisotropy easy-axis, macrospin ferromagnets where the exchange stiffness is large enough to preclude any inhomogeneous textures, such as domain walls or spin spirals. Moreover, we do not consider the Dzyaloshinskii–Moryia-type[30] interactions at the interfaces, which may lead to noncollinear magnetization configurations. Therefore, the free energy of such a system is defined by

$$F = H_0 - \frac{1}{\beta} \sum_n \ln(1 + e^{-\beta E_n / 2}). \quad (1)$$

Here, $\beta = \frac{1}{k_B T}$ and $k_B$ is the Boltzmann constant and $T$ is the temperature. $H_0$ is a constant term to be specified later, which consists of a superconducting constant term ($H_0^S$) and chemical potential constant term ($H_0^\mu$). $H_0^S$ arises as a result of performing a mean-field approximation while $H_0^\mu$ is due to a symmetrization of the Hamiltonian. Moreover, $E_n$ is the $n$th eigenvalue and will be calculated by means of diagonalizing a tight-binding Hamiltonian for the structure of interest. The Hamiltonian is as follows,

$$H = -\sum_{\langle i,j \rangle,\alpha} t_{ij} c_{i\alpha}^\dagger c_{j\alpha} - \sum_{i,\alpha} \mu_i n_{i\alpha} - \sum_i U_i n_{i\uparrow} n_{i\downarrow} - \sum_{i\alpha\beta} (\mathbf{h}_i^\delta \cdot \boldsymbol{\sigma})_{\alpha\beta} c_{i\alpha}^\dagger c_{i\beta} \, . \tag{2}$$

Here, $c_{i\alpha}^\dagger$ ($c_{i\alpha}$) creates (annihilates) an electron with spin $\alpha$ at site $i = (i_x, i_y)$ with $i_x = 1, \cdots N_x$ and $i_y = 1, \cdots N_y$. Also, $t_{ij}$ is the hopping integral between nearest-neighbor sites and we assume that it has a constant value $t$. $\mu_i$ is the chemical potential at site $i$ while $n_{i\alpha} = c_{i\alpha}^\dagger c_{i\alpha}$ is the number operator. The fourth term in the Hamiltonian represents the local exchange interaction with $h_i^\delta$ being the strength of this field in the left ($\delta = L$) or right ($\delta = R$) ferromagnets. Also, $h_i^L = h_i(0, 1, 0)$, $h_i^R = h_i(\sin(\theta), \cos(\theta), 0)$ and $\boldsymbol{\sigma} = (\sigma_x, \sigma_y, \sigma_z)$ the Pauli matrices. Although considering $h$ as a fixed input parameter is a standard approach in much of the literature, we have also checked if our results hold when the magnetization is solved self-consistently as well. A brief study of $J$ with self-consistent magnetization is included in the Supplementary information. We find that solving the magnetization self-consistently has very little effect on the results for the case when we have open boundary conditions along the $y$-direction. It does not change the physics for the 2D system compared to using a fixed input-value for $h$. We consider a singlet superconductor for the central part, modelling the interaction as an on-site attractive $U$ as the third term of Hamiltonian. $U_i = U > 0$ is the local attractive interaction which creates Cooper pairs in the superconductor while it is zero elsewhere. We treat the interaction term by a mean-field approximation to simplify the problem,

$$-\sum_i U_i n_{i\uparrow} n_{i\downarrow} = -\sum_i U_i \left( c_{i\uparrow}^\dagger c_{i\downarrow}^\dagger \langle c_{i\downarrow} c_{i\uparrow} \rangle + c_{i\downarrow} c_{i\uparrow} \langle c_{i\uparrow}^\dagger c_{i\downarrow}^\dagger \rangle - \langle c_{i\downarrow} c_{i\uparrow} \rangle \langle c_{i\uparrow}^\dagger c_{i\downarrow}^\dagger \rangle \right). \tag{3}$$

If we define superconducting gap as $\Delta_i = -U_i \langle c_{i\downarrow} c_{i\uparrow} \rangle$, then

$$-\sum_i U_i n_{i\uparrow} n_{i\downarrow} = \sum_i (c_{i\uparrow}^\dagger c_{i\downarrow}^\dagger \Delta_i + c_{i\downarrow} c_{i\uparrow} \Delta_i^*) + H_0^S, \tag{4}$$

where we have defined

$$H_0^S = \sum_i \frac{|\Delta_i|^2}{U_i} \, . \tag{5}$$

The Hamiltonian does not contain any constant mean-field term containing the magnetic order parameter since we do not solve for the magnetization self-consistently.

We proceed to explain how the eigenvalues $E_n$ are obtained. Our Hamiltonian Eq. (2) is bilinear in the fermion operators and can be diagonalized. Choosing the following basis,

$$W^\dagger = \begin{bmatrix} D_1^\dagger & D_2^\dagger & D_3^\dagger & \ldots & D_{N_y}^\dagger \end{bmatrix}, \tag{6}$$

where we have defined

$$D_{i_y}^\dagger = \begin{bmatrix} B_{(1,i_y)}^\dagger & B_{(2,i_y)}^\dagger & B_{(3,i_y)}^\dagger & \ldots & B_{(N_x,i_y)}^\dagger \end{bmatrix} \tag{7}$$

with $i_y = 1, \cdots N_y$ and

$$B_i^\dagger = \begin{bmatrix} c_{i\uparrow}^\dagger & c_{i\downarrow}^\dagger & c_{i\uparrow} & c_{i\downarrow} \end{bmatrix}, \tag{8}$$

the Hamiltonian may now be written as

$$H = H_0 + \frac{1}{2} W^\dagger S W = H_0 + \frac{1}{2} \sum_{ij} B_i^\dagger h_{ij} B_j \, . \tag{9}$$

Here, $H_0$ is the constant term that we discussed previously, and

$$S = \begin{bmatrix} S_{11} & \cdots & S_{1,N_y} \\ \vdots & \ddots & \vdots \\ S_{N_y,1} & \cdots & S_{N_y,N_y} \end{bmatrix} \tag{10}$$

with

$$S_{i_y,j_y} = \begin{bmatrix} h_{(1,i_y)(1,j_y)} & \cdots & h_{(1,i_y)(N_x,j_y)} \\ \vdots & \ddots & \vdots \\ h_{(N_x,i_y)(1,j_y)} & \cdots & h_{(N_x,i_y)(N_x,j_y)} \end{bmatrix}. \tag{11}$$

Finally, the $4 \times 4$ matrix for interaction between sites $i$ and $j$ is

$$h_{ij} = -\left[\frac{t}{2}(\delta_{i_x,j_x-1} + \delta_{i_x,j_x+1}) + \mu_i \delta_{i_x,j_x}\right]\delta_{i_y,j_y}\tau_3\sigma_0 - \left[\frac{t}{2}(\delta_{i_y,j_y-1} + \delta_{i_y,j_y+1}) + \mu_i \delta_{i_y,j_y}\right]\delta_{i_x,j_x}\tau_3\sigma_0 + \left[-h_i^z\tau_3\sigma_z - h_i^x\tau_3\sigma_x - h_i^y\tau_0\sigma_y\right.$$
$$\left. + \Delta_{i_x,i_y} i\tau^+\sigma_y - \Delta^*_{i_x,i_y} i\tau^-\sigma_y\right]\delta_{i_x,j_x}\delta_{i_y,j_y}.$$
(12)

Here, $\tau_m\sigma_l = \tau_m \otimes \sigma_l$ and $\tau^\pm = \frac{1}{2}(\tau_1 \pm i\tau_2)$. $S$ is Hermitian and can be diagonalized numerically. Note that we are considering a finite size 2D system without any periodic boundary conditions. The 2D model is an approximation that is necessary because doing the calculations in 3D becomes numerically too demanding in terms of computational time. Although it could be interesting to consider how a 3D computation alters the result, we do not expect any qualitatively new effects. Diagonalizing the Hamiltonian by introducing a new basis gives

$$H = H_0 + \frac{1}{2}\sum_n E_n \gamma_n^\dagger \gamma_n.$$
(13)

The eigenfunctions for $S$ are

$$\Phi_n^\dagger = \left[\phi_{1n}^\dagger \ \phi_{2n}^\dagger \ \cdots \ \phi_{N_y,n}^\dagger\right],$$
(14)

where we have defined

$$\phi_{i_y,n}^\dagger = \left[\varphi_{(1,i_y),n}^\dagger \ \varphi_{(2,i_y),n}^\dagger \ \cdots \ \varphi_{(N_x,i_y),n}^\dagger\right],$$
$$\varphi_{(i_x,i_y)n}^\dagger = \left[v_{(i_x,i_y),n}^* \ v_{(i_x,i_y),n}^* \ \omega_{(i_x,i_y),n}^* \ \chi_{(i_x,i_y),n}^*\right].$$
(15)

The original creation and annihilation operators $\{c^\dagger, c\}$ now can be expressed with new quasiparticle operators,

$$c_{i\uparrow} = \sum_n v_{i,n}\gamma_n, \ c_{i\downarrow} = \sum_n v_{i,n}\gamma_n,$$
$$c_{i\uparrow}^\dagger = \sum_n \omega_{i,n}\gamma_n, \ c_{i\downarrow}^\dagger = \sum_n \chi_{i,n}\gamma_n.$$
(16)

Using these, we obtain a self-consistency equation for $\Delta_i$,

$$\Delta_i = -U_i \sum_n v_{i,n}\omega_{i,n}^*(1 - f(E_n/2)).$$
(17)

The local density of states (LDOS) is the density of states at one site and in our model it can be calculated for $T = 0$. The number of charges at site $i$ is given by

$$\rho_i = \sum_\alpha \langle c_{i\alpha}^\dagger c_{i\alpha}\rangle.$$
(18)

At an arbitrary temperature, the number of charges at site $i$ is

$$\rho_i = \int_{-\infty}^{+\infty} N_i(E)f(E)dE.$$
(19)

Here, $N_i(E)$ is the local density of states at site $i$ and $f(E)$ is the Fermi-Dirac distribution with energy $E$ measured relative the chemical potential. When $T = 0$, we know that $f(E) = 1$ for $E < 0$ and $f(E) = 0$ when $E > 0$. Therefore, the LDOS takes the form:

$$N_i(E) = \sum_n (|v_{in}|^2 + |v_{in}|^2)\delta(E_n/2 - E).$$
(20)

In our model, the proximity effect of the superconductor into the ferromagnets (induced Cooper pair correlations) is quantified by the anomalous Green function $F_i = -\langle c_{i\downarrow}c_{i\uparrow}\rangle$. Also, the inverse proximity effect causing an induction of magnetic polarization in the superconductor is accounted for by $M_i^y = \langle S_i^y \rangle$ where the $\langle\ldots\rangle$ notation denotes expectation value. The spin operator is $S_i^y = \sum_{\alpha\beta} c_{i\alpha}^\dagger(\sigma_{\alpha\beta})_y c_{i\beta}$. Therefore, the magnetization along the $y$-direction in the system is given by

$$M_i^y = \sum_n i(-v_{i,n}^* v_{i,n} + v_{i,n}^* v_{i,n})f\left(\frac{E_n}{2}\right)$$
(21)

Equation (17) is solved self-consistently. By considering an initial value for the gap, we diagonalize the Hamiltonian to obtain the eigenvalues ($E_n$), and eigenfunctions [Eq. (14)] to compute $v_{i,n}$ and $\omega_{i,n}$. By means of Eq. (17), we can then compute a new value for the gap and once again diagonalize the Hamiltonian with the new value for the gap. This process is repeated until the relative change in the gap value between iterations is smaller than the convergence criterion. The criterion for numerical convergence of the superconducting gap is set to be a relative change of $5 \times 10^{-4}$. To ensure that the self-consistent solution converges to the ground-state, we have checked several possible initial values for the superconducting order parameter.

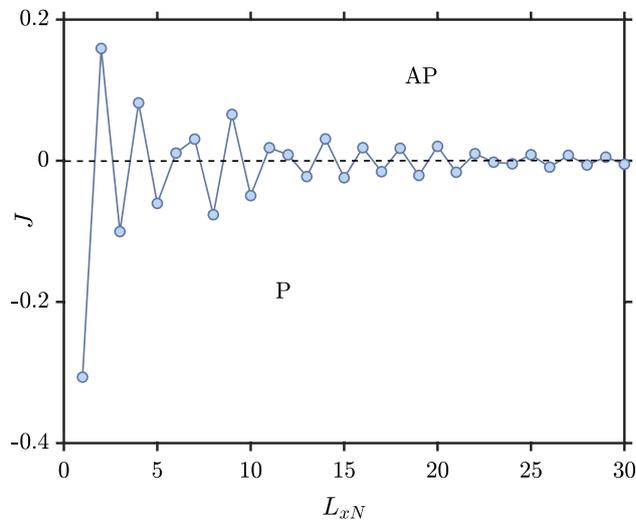

**Figure 2.** The indirect exchange interaction $J$ between the two ferromagnetic contacts mediated by a normal material (F–N–F structure) when $L_y = 10, L_{xF} = 2, \mu_N = 0.8t, \mu_F = 0.9t, |h_i^L| = |h_i^R| = h_i = 1t$ and $k_B T = 0.01t$.

## Results

**F–N–F junction, briefly revisited.** The main purpose of this paper is to investigate the indirect exchange coupling between two ferromagnets separated by a superconductor when solving self-consistently for the order parameter and taking into account both the proximity effect and the inverse proximity effect. We will consider a large range of $h$-values corresponding to either weakly polarized ferromagnets, such as PdNi, or strongly polarized elemental ferromagnets like Co or Fe. Before considering the superconducting case, it is worth considering briefly a three layer F–N–F structure as shown in Fig. 1. We include this treatment so that the reader can more easily contrast the normal and superconducting case. We choose a representative set of parameters as $L_y = 10$, $L_{xF} = 2$, $\mu_N = 0.8t$, $\mu_F = 0.9t$, and $k_B T = 0.01t$. In our notation, $L_y$ is the number of lattice sites along the $y$-axis and $L_{xF}$ is the number of lattice sites along the $x$-axis in the ferromagnets. The length of the ferromagnetic part has little influence on the final results in the F–N–F case and also does not change the results qualitatively in the F–S–F case. Therefore, we have chosen a small value for $L_{xF}$ to reduce the required time of the numerical simulations. Both ferromagnetic contacts have the same exchange field strength and the magnetization is directed along $\hat{y}$ ($|h_i^L|=|h_i^R|=h_i$). As the length of the normal part increases, the amplitude of the well-known RKKY-like oscillations in the F–N–F structure decreases as shown in Fig. 2. These oscillations indicate a switching between P and AP configurations as the ground-state of the junction: $J > 0$ corresponds to an AP configuration, while $J < 0$ corresponds to a P configuration.

Figure 3 shows $J$ as a function of the exchange field strength in the ferromagnets ($h_i$) for several different normal region lengths. It demonstrates that $J$ not only oscillates as a function of $L_{xN}$, but also as a function of $h_i$. The oscillations stem from the fact that the eigenstates for the quasiparticle excitations in the system interfere constructively or destructively at the ferromagnetic contacts, depending on the length $L_{xN}$ and the exchange field $h_i$ since both these quantities determine the phase-change of an eigenstate as one moves across the normal metal.

Despite the oscillations for small exchange field strengths, $J$ monotonically decreases when $h_i$ becomes sufficiently large. This decay is likely related to the depletion in the number of available states around the Fermi level in the ferromagnetic part as shown in Fig. 4.

**F–S–F junction.** We now turn to the main topic of this manuscript, namely a study of how the exchange interaction between two ferromagnets is mediated by an *s*-wave superconductor. Our results for the RKKY interaction in the superconducting state will be relevant for materials which have a (nearly) isotropic superconducting gap and where the ratio $\Delta/\mu$ between the gap $\Delta$ and the Fermi energy $\mu$ is relatively large. The reason for the latter criterion is that large values of the gap is required in the lattice BdG formalism in order to work with systems that have a computationally manageable size. Although our model is simplified and approximative, there exists materials which matches the parameter-choice in our manuscript for the energy scales involved. One such superconducting material is FeSe[31] which is known to have a large ratio $\Delta/\mu$ of order $\sim 0.1$, as in our parameter choice. The exact crystal structure of FeSe differs from our model and its full tight-binding model is more complex than the simple model used in our manuscript, but the projection of the FeSe crystal structure onto the plane is in fact a square lattice. Another material with a high ratio $\Delta/\mu$ is FeTe$_{0.6}$Se$_{0.4}$[32] which would also resemble our parameter-choice in terms of the relative size of the gap and Fermi energy.

In Fig. 5, we plot $J$ against the length of the superconducting region for three different on-site pairing interactions $U/t = 1, U/t = 1.5$ and $U/t = 2$. The superconducting gap $\Delta$ tends to zero for a short superconductor with a weak superconducting interaction $U$. For the case of $U/t = 1$, $J$ is the same as the F–N–F case. This is simply because $\Delta$ is zero for short superconductors and when $L_{xS}$ has increased sufficiently to render $\Delta$ non-zero, $J \to 0$ since the distance between the ferromagnets is then too large.

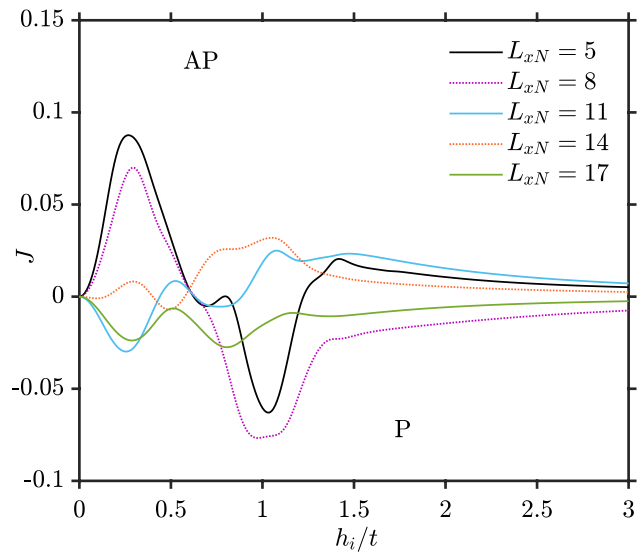

**Figure 3.** Indirect exchange interaction as a function of the exchange field strength of the ferromagnets for the F–N–F structure). Here, $L_y = 10, L_{xF} = 2, \mu_N = 0.8t, \mu_F = 0.9t$ and $k_B T = 0.01t$.

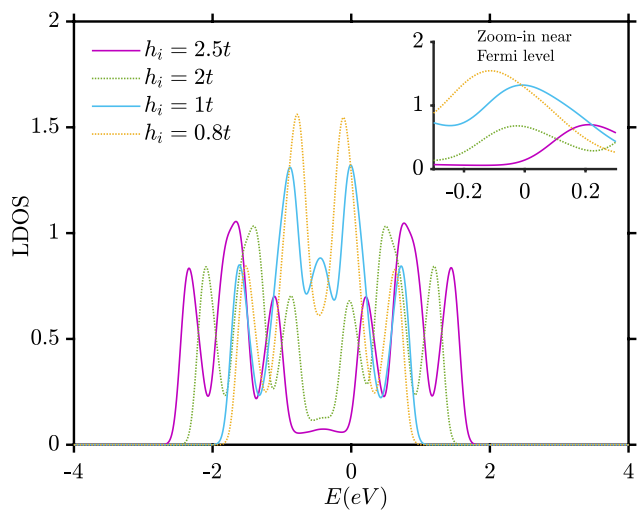

**Figure 4.** Local density of states (LDOS) for the (2,2) site inside the left ferromagnet as a function of energy for 4 different $h_i$. Here, $L_y = 10, L_{xF} = 2, L_{xS} = 8, \mu_N = 0.8t, \mu_F = 0.9t, k_B T = 0.01t$. The dashed box indicates an area around the Fermi energy.

For the case of $U/t = 1.5$, there are two mechanisms competing against each other. One is the conventional RKKY-like oscillations mediated by quasiparticles. The other mechanism is the blocking of states that can mediate the interaction due to the superconducting gap. This can be seen from blue curve of Fig. 5. For short superconductors, the RKKY-like oscillations are approximately the same as in the F–N–F case because the gap is too small to block any significant fraction of the quasiparticles. For longer superconductors the gap increases and dominates the indirect exchange interaction $J$.

Figure 6a,b show that for $L_{xS} = 4$ and $L_{xS} = 5$, the gap is finite in both the P and AP configuration, but still RKKY-like oscillations dominate as seen in Fig. 5 for $U/t = 1.5$. However, as $L_{xS}$ increases in Fig. 6c,d, $\Delta$ becomes sufficiently large to block the oscillations caused by quasiparticles. Now, we see that $\Delta_P > \Delta_{AP}$ which leads to $J > 0$, favoring an AP magnetic configuration as the ground state. At first glance, this might seem strange since a larger $\Delta$ in the P configuration should give a larger superconducting condensation energy gain compared to the AP configuration. However, the configuration with the largest gap will also block the largest amount of quasiparticles that can mediate the interaction between the ferromagnets and lower the free energy. In our numerical simulations, we find that when the gap is large enough in magnitude, it is the latter blocking effect that determines the ground-state of the system. Hence, $\Delta_P > \Delta_{AP}$ causes $J > 0$.

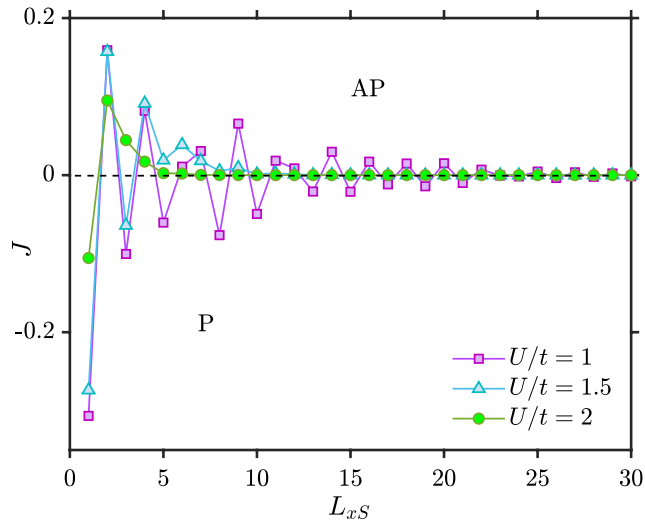

**Figure 5.** $J$ vs the length of superconducting part ($L_{xS}$) for the F–S–F structure. When $L_y = 10$, $L_{xF} = 2$, $\mu_S = 0.8t, \mu_F = 0.9t, k_B T = 0.01t, |h_y^L| = |h_y^R| = h_i = 1t$.

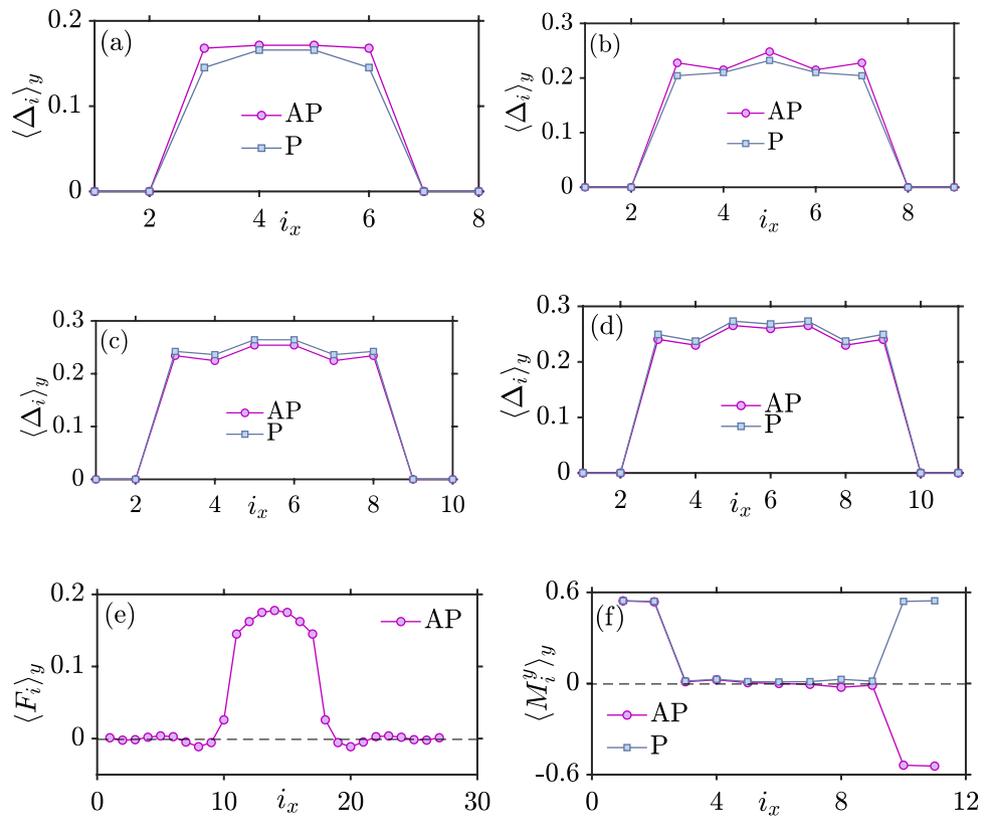

**Figure 6.** AP and P superconducting gaps ($\Delta_{AP}$ and $\Delta_P$) when $U/t = 1.5, h_i = 1t, k_B T = 0.01t, \mu_S = 0.8t$ and $\mu_F = 0.9t$ (**a**) $L_{xS} = 4$ (**b**) $L_{xS} = 5$ (**c**) $L_{xS} = 6$ (**d**) $L_{xS} = 7$. (**e**) Superconducting proximity effect inside the ferromagnets when $U/t = 1.5, h_i = 0.5t, k_B T = 0.01t, \mu_S = 0.8t, \mu_F = 0.9t, L_{xS} = 7$ and $L_{xF} = 10$. (**f**) Layer-dependent magnetization for P and AP when $U/t = 1.5, h_i = 1t, k_B T = 0.01t, \mu_S = 0.8t, \mu_F = 0.9t$, $L_{xS} = 7$ and $L_{xF} = 2$. The $\langle \ldots \rangle_y$ notation denotes averaging over the $y$-direction and $i$ is a lattice site along the $x$-direction.

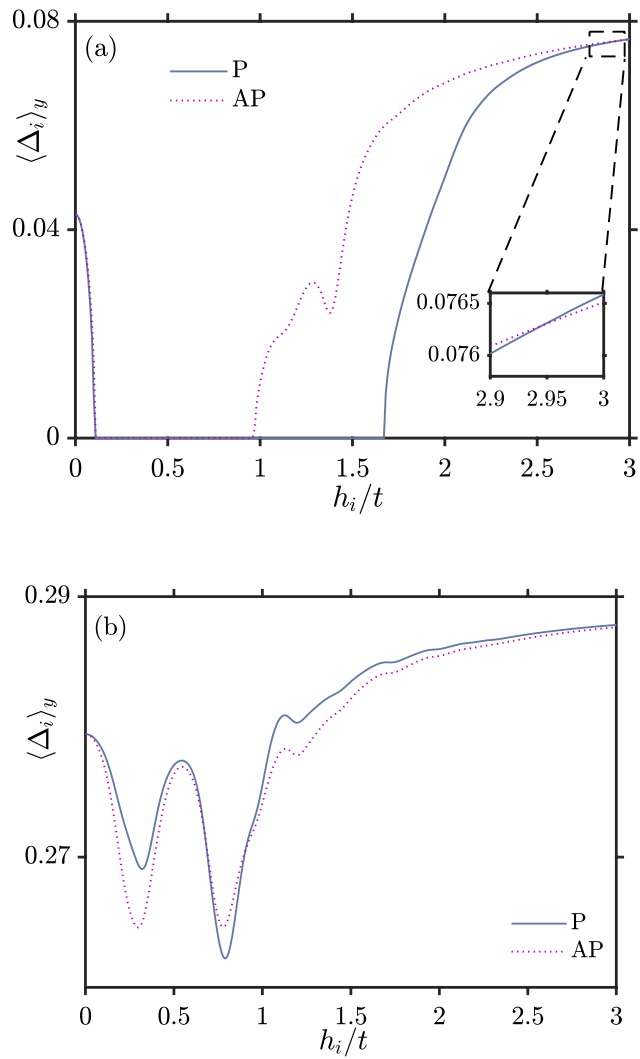

**Figure 7.** Superconducting gap vs $h_i$ when $L_{xS} = 8$, $k_B T = 0.01t$, $\mu_S = 0.8t$ and $\mu_F = 0.9t$. (**a**) $U/t = 1$ (**b**) $U/t = 1.5$.

In Fig. 6, we demonstrate the presence of a superconducting proximity effect inside the ferromagnets. Whereas the gap $\Delta_i$ vanishes inside the ferromagnets due to the absence of an attractive interaction $U_i$ in those regions in our model, the anomalous Green function $\langle -c_{i\downarrow} c_{i\uparrow} \rangle$ is finite in the ferromagnets, as shown in Fig. 6e. In that plot, we have considered larger ferromagnets ($L_{xF} = 10$) so that the oscillatory nature of the Cooper pairs penetrating inside the ferromagnets is better shown. On the other hand, an inverse proximity effect is also present: the induction of a magnetization induced inside the superconductor due to the exchange field of the ferromagnets. The layer-dependent magnetization is shown in Fig. 6 (f).

It is often assumed in the literature that the AP configuration in a superconducting spin valve should give the largest superconducting gap. The rationale behind this assumption is that the induced magnetization in the superconducting region of the F–S–F structure is weakest in the AP configuration, leading to the the least amount of pair-breaking. However, as we will discuss below, this is a simplified picture which neglects a key process in the spin valve: crossed Andreev reflection. The effect on $\Delta$ of various pair-breaking processes in equilibrium F–S–F structures has been studied previously, but primarily in layers with monoatomic thickness[33–38]. In Ref.[37], it was stated that $\Delta_P < \Delta_{AP}$ at any temperature for sufficiently large thicknesses. In our work, we instead find that the opposite inequality holds for sufficiently large thicknesses of the superconductor.

For $U/t = 2$ in Fig. 5 one observes a monotonic decrease of $J$ as a function of $L_{xS}$. This behavior occurs both for a strong pairing interaction $U$ or when the exchange field $h_i$ is large. We have already explained why it occurs for strong $U$, leading to a large gap. To explain why it occurs for a large exchange field, we consider the behaviour of $\Delta$ as a function of exchange field strength: this is shown in Fig. 7a for $U/t = 1$ and Fig. 7b for $U/t = 1.5$.

In both cases, for large enough $h_i$, there exists a specific $h_i$ value which marks the transition from $\Delta_P < \Delta_{AP}$ to $\Delta_P > \Delta_{AP}$. The reason for this transition can be explained in terms of a competition between the pair-breaking influence of the induced exchange field in the superconductor and inverse crossed Andreev reflection (CAR)[39], which we proceed to explain.

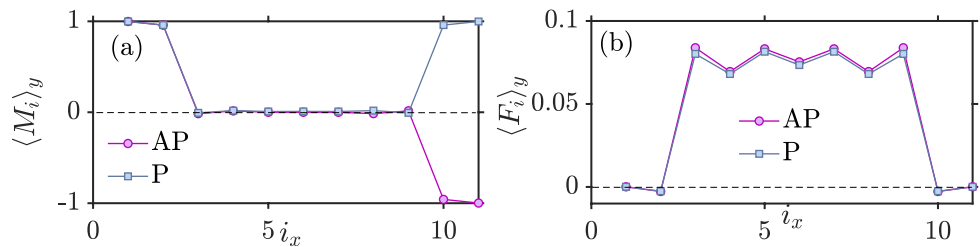

**Figure 8.** (**a**) Layer-dependent magnetization for P and AP (**b**) Proximity effect of superconductor inside the ferromagnets when $U/t = 1, h_i = 2.5t, k_B T = 0.01t, \mu_S = 0.8t, \mu_F = 0.9t$ and $L_{xS} = 7$.

In a superconducting spin valve, a magnetization is induced inside the superconductor. The corresponding induced exchange field is stronger in the case of P orientation than the AP one. As a result of this induced exchange field, the opposite spin electrons in the Cooper pair accumulate different phases in the superconductor, ultimately leading to a loss of phase coherence and Cooper-pair breaking. As the induced exchange field is stronger in the P case, the destructive effect of the ferromagnets is more severe in the P orientation than the AP. If this was the only mechanism, the superconducting gap should always be smaller in the P orientation ($\Delta_P < \Delta_{AP}$).

On the other hand, inverse crossed Andreev reflection is another pair breaking mechanism in competition with the pair breaking effect of the induced exchange field. What happens here is that spin up and down electrons in a Cooper pair move into separate ferromagnets. This is in contrast to the usual proximity effect mediated by local Andreev reflection, where both electrons (and thus the entire pair) leak into a single material. Crossed Andreev reflection is thus a non-local process. In the AP orientation, the electrons tunnel into the spin majority band of the two ferromagnets while in the P orientation one spin goes to a majority band while the other one goes to the minority band of the other ferromagnet.

As the exchange field becomes stronger, CAR becomes less probable to occur in the P configuration since the minority band involved in the process gradually vanishes. In the half-metallic limit, there is no longer any conducting minority band to enable CAR in the P configuration. Therefore, the destructive effect of CAR is stronger in the AP case, thus making the gap smaller ($\Delta_{AP} < \Delta_P$).

The configuration giving the largest $\Delta$ then depends on which of the two described effects that dominates. From Fig. 7, the fact that $\Delta_P$ overtakes $\Delta_{AP}$ in magnitude at a critical value for the exchange field $h_i$ indicates that crossed Andreev reflection dominates in this regime. This reduces the leakage of superconductivity into the ferromagnets, and enhances the gap. It can be done by means of rotating the magnetization direction of one of the ferromagnets, for example with the help of an external magnetic field.

Figure 7a also demonstrates an example of reentrant superconductivity which can occur in superconducting spin valves, enabling a switching of superconductivity on and off possible. This is usually done by varying thickness of one of the ferromagnets in a superconducting spin valve[14,40]. The suppression of superconductivity by the ferromagnet becomes particularly effective at certain ferromagnet thicknesses $L_F$. At these values of $L_F$, the different quasiparticle trajectories inside the ferromagnet interfere in such a manner that they minimizes the superconducting condensate wave function at the interface with the superconductor. The interference occurs since the quasiparticles amplitudes are quantum mechanically determined from sum over all classical trajectories. Since phases picked up along these trajectories not only depend on the length of the trajectory, but also the magnitude of the exchange field, varying $h$ also leads to reentrant behavior of superconductivity in our case.

Another interesting aspect of this plot (Fig. 7a) is that superconductivity is eventually enhanced when $h_i/t$ increases compared to the case $h_i = 0$ without magnetization. In other words, strong magnetization enhances superconductivity. The reason for this effect can be understood by considering the mechanisms causing a suppression of superconductivity: local Andreev reflection occurring at a single interface in the system, crossed Andreev-reflection (CAR) occurring at both interfaces in the system, and the pair-breaking magnetic moment induced in the superconducting region when the exchange-field is finite. In the case of no exchange field ($h_i = 0$), the two first mechanisms listed above are then at play and reduce the superconducting gap. When exchange field is strong, the gap is seen to be enhanced compared to the $h_i = 0$ case in Fig. 7. Moreover, as seen from Fig. 7, the gap becomes almost identical in both the P and AP configuration. This indicates that the reason for the enhancement compared to $h_i = 0$ cannot be CAR or the induced magnetic moment (Fig. 8a), because these two mechanisms act very differently in the P and AP configuration. Instead, the reason is the first mechanism listed above: the behavior of local Andreev reflection. Namely, as the exchange field grows in magnitude, the ferromagnets become closer to being half-metallic (only conducting in one spin-band). For a half-metal/superconductor interface, there is in fact no proximity effect at all (in Fig. 8b, the proximity effect goes to zero quickly in the ferromagnets). As only one spin-type is available in the half-metal, no Andreev reflection can take place and the Cooper pairs are confined to the superconductor. Therefore, unlike the case for $h_i = 0$, there is now no leakage of Cooper pairs into the ferromagnetic regions. This results in a stronger superconducting condensate, since the leakage of pairs into the magnetic regions is absent.

In Fig. 9, we show how $J$ in a superconducting spin valve behaves with respect to exchange field. The interaction between the ferromagnets weakens the longer the superconductor is. Despite of a small region where the P orientation is the ground state, it is clearly seen that AP is mostly the dominating ground state, especially as $h_i$ becomes large. As we mentioned previously, this is as a result of $\Delta_P$ exceeding the magnitude of $\Delta_{AP}$. Similarly to the F–N–F case, for high enough exchange field $h_i$ the number of available conduction electron states near the Fermi level that can become spin-polarized and mediate the interaction monotonically decreases, leading to a corresponding reduction of the indirect exchange interaction.

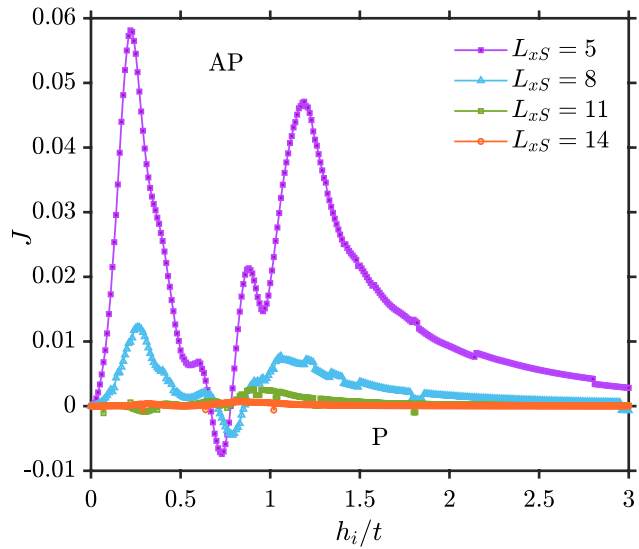

**Figure 9.** F–S–F, $L_y = 10, L_{xF} = 2, \mu_S = 0.8t, \mu_F = 0.9t, k_B T = 0.01t, U = 1.5$.

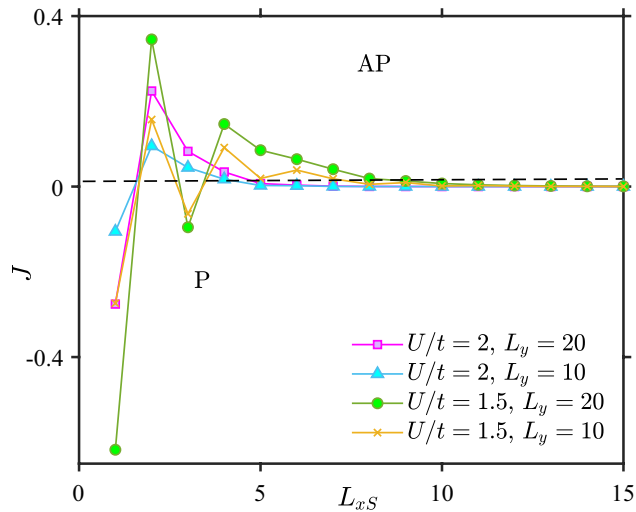

**Figure 10.** F–S–F, $L_{xF} = 2, \mu_S = 0.8t, \mu_F = 0.9t, k_B T = 0.01t$.

In Ref.[41], the authors showed that the critical temperature difference $\Delta T_c \equiv T_c(P) - T_c(AP)$ in a superconducting spin-valve could have both positive and negative sign if the thicknesses of the ferromagnets were unequal. The authors considered very thin superconductors $L_S \ll \xi_S$. In our manuscript, we do not consider such thin superconductors and instead focus on the regime $L_S > \xi_S$. Moreover, Ref.[41] obtained the sign change in $\Delta T_c$ when assuming a strong Fermi-vector mismatch between the F and S regions, whereas we do not consider layers with a strong Fermi-vector mismatch. In our regime, we do not observe any sign-change in the indirect exchange interaction $J$ as we vary the thickness of one of the ferromagnets while keeping the other fixed. This indicates that the interference effects causing the sign change in $T_c(P) - T_c(AP)$ in Ref.[41] are suppressed for superconductors larger than the coherence length and when there is no strong Fermi-vector mismatch between the layers.

The relevant length-scales under consideration in our system are the superconducting coherence length $\xi = \hbar v_F / \pi \Delta$, the ferromagnetic coherence length $\xi_F = \hbar v_F / \pi h$, and the Fermi wavelength $\lambda_F = k_F^{-1}$. We have considered several different parameter choices for the length of the superconductor and the coherence length (by varying $U$). Therefore, the relative size of these length-scales is not a fixed number in our paper. However, considering a representative parameter set $L_S/a = 10, U/t = 1.5, h/t = 1.0$, we find that $\xi_S/\lambda_F \simeq 3.5, \xi_F/\lambda_F \simeq 1.2$. In this work, we have not considered the possible influence of the electromagnetic proximity effect[42,43] on the RKKY interaction between the ferromagnets, which could be an interesting extension to consider.

To close this section, we investigate the effect of the width of the structure on the indirect interaction between the ferromagnets in Fig. 10. Increasing the width of the structure leads to more available states which makes the interaction $J$ between the ferromagnets larger. However, it does not change the fact that the system prefers the AP oreintation as the ground state for sufficiently strong superconductors.

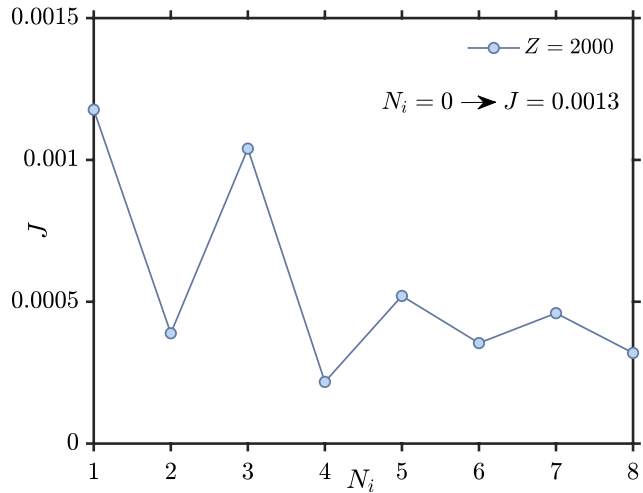

**Figure 11.** Indirect exchange interaction $J$ between the ferromagnets versus number of impurities ($N_i$). We have considered $J$ averaged over 2000 different impurity configurations, using $L_y = 10, L_{xF} = 2, L_{xS} = 10, \mu_S = 0.8t, \mu_F = 0.9t, k_B T = 0.01t, U/t = 1.5$ and $h_i = 1t$.

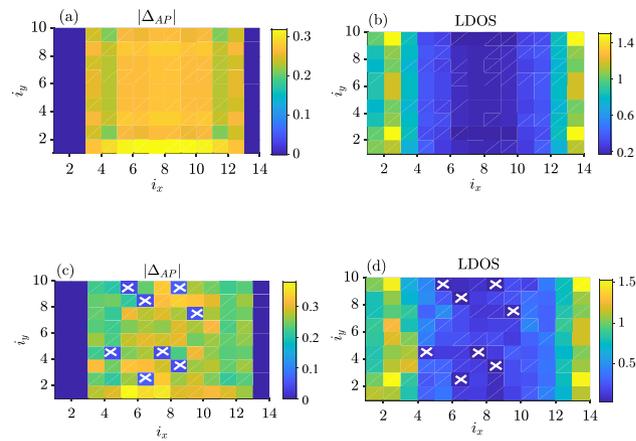

**Figure 12.** Left column: anti-parallel superconducting gap ($|\Delta_{AP}|$). Right column: local density of states (LDOS) plots. We have used $L_{xS} = 10, L_y = 10, L_{xF} = 2, \mu_S = 0.8t, \mu_F = 0.9t, U/t = 1.5, h_i = 1t$ and $\Phi = 0.036$. (**a**, **b**) clean F–S–F ($N_i = 0$). (**c**, **d**) dirty F–S–F with 8% impurity concentration ($N_i = 8$).

We finally consider the effect of impurities on $J$. To this end, we consider randomly located impurities in the superconducting part. Impurity atoms are not chosen from edge atoms and atoms at the interfaces with ferromagnets. Here, we consider the impurity-averaged exchange interaction $J$ over a large set of different impurity configurations. We define $Z$ as the number of impurity configurations that we have averaged over. The Hamiltonian of the system including impurity scattering is as follows

$$H = -\sum_{\langle ij \rangle, \alpha} t_{ij} c_{i\alpha}^\dagger c_{j\alpha} + \sum_{i,\alpha} (V_i^{\text{imp}} - \mu_i) n_{i\alpha} + \sum_{i\alpha\beta} (\mathbf{h}_i \cdot \boldsymbol{\sigma})_{\alpha\beta} c_{i\alpha}^\dagger c_{i\beta} - \sum_i U_i n_{i\uparrow} n_{i\downarrow}. \tag{22}$$

Here $V_i^{\text{imp}}$ is the potential describing the impurity strength at site $i$. In Fig. 11, we consider $J$ as a function of the number of impurities in the system for $U/t = 1.5, h_i = 1t$ and $V_i^{\text{imp}} = 2t$, averaging over $Z = 2000$ configurations. We see that $J$ decays in an oscillatory fashion as the number of impurities randomly placed in the superconductor increases.

To understand the behavior of $J$, we consider both how the magnitude and the LDOS changes for the F–S–F structure when comparing the clean case and the case with impurities. Consider first the case with zero impurities and zero magnetic field, shown in Fig. 12a,b. The LDOS has its minimum value in the middle of structure while $|\Delta_{AP}|$ is maximal at the middle of structure, as expected. When adding impurities, in Fig. 12c,d, $|\Delta_{AP}|$ will tend to zero around the impurity atoms. Their location is marked with white crosses. Interestingly, the average LDOS in the dirty F–S–F case (Fig. 12d) has increased in comparison to the clean F–S–F (Fig. 12b) case. At first glance, this might indicate that more available quasiparticle states are available to mediate the exchange interaction between the ferromagnets. This should lead to an increase in $J$ compared to the clean case $N_i = 0$. However, Fig. 11 shows the opposite: $J$ is reduced compared to the clean case. We attribute this decrease in $J$ with increasing impurity concentration to an increasing localization of quasiparticles[28,29]. When the localization increases, the interaction $J$ should be reduced, as seen in Fig. 11.

## Concluding remarks

In conclusion, we have considered the magnetic exchange interaction $J$ and the preferred equilibrium magnetic configuration in a 2D superconducting spin valve with an $s$-wave superconductor, solving self-consistently for the superconducting order parameter. We find that the qualitative dependence of $J$ on the separation distance between the ferromagnets can behave differently on the basis of the strength of the superconducting gap and the strength of the exchange field in the ferromagnets. RKKY-like oscillations are observed when the superconducting gap $\Delta$ is small, whereas a monotonic decay is observed when $\Delta$ is larger. In the latter case, the AP configuration is always preferred even though the gap is larger in the P configuration. We explain this in terms of a competition between a proximity-induced pair-breaking magnetization in the superconductor and crossed Andreev reflection. Adding randomly localized impurities to the superconductor led to an oscillatory decrease of $J$ with increasing impurity concentration.

## References

1. Ruderman, M. A. & Kittel, C. Indirect exchange coupling of nuclear magnetic moments by conduction electrons. *Phys. Rev.* **96**, 99–102. https://doi.org/10.1103/PhysRev.96.99 (1954).
2. Baibich, M. N. *et al.* Giant magnetoresistance of (001)fe/(001)cr magnetic superlattices. *Phys. Rev. Lett.* **61**, 2472–2475. https://doi.org/10.1103/PhysRevLett.61.2472 (1988).
3. Binasch, G., Grünberg, P., Saurenbach, F. & Zinn, W. Enhanced magnetoresistance in layered magnetic structures with antiferromagnetic interlayer exchange. *Phys. Rev. B* **39**, 4828–4830. https://doi.org/10.1103/PhysRevB.39.4828 (1989).
4. Zhu, J.-J., Yao, D.-X., Zhang, S.-C. & Chang, K. Electrically controllable surface magnetism on the surface of topological insulators. *Phys. Rev. Lett.* **106**, 097201. https://doi.org/10.1103/PhysRevLett.106.097201 (2011).
5. Abanin, D. A. & Pesin, D. A. Ordering of magnetic impurities and tunable electronic properties of topological insulators. *Phys. Rev. Lett.* **106**, 136802. https://doi.org/10.1103/PhysRevLett.106.136802 (2011).
6. Sherafati, M. & Satpathy, S. Rkky interaction in graphene from the lattice green's function. *Phys. Rev. B* **83**, 165425. https://doi.org/10.1103/PhysRevB.83.165425 (2011).
7. Hosseini, M. V. & Askari, M. Ruderman–Kittel–Kasuya–Yosida interaction in Weyl semimetals. *Phys. Rev. B* **92**, 224435. https://doi.org/10.1103/PhysRevB.92.224435 (2015).
8. Liu, Q., Liu, C.-X., Xu, C., Qi, X.-L. & Zhang, S.-C. Magnetic impurities on the surface of a topological insulator. *Phys. Rev. Lett.* **102**, 156603. https://doi.org/10.1103/PhysRevLett.102.156603 (2009).
9. Chesi, S. & Loss, D. Rkky interaction in a disordered two-dimensional electron gas with Rashba and Dresselhaus spin-orbit couplings. *Phys. Rev. B* **82**, 165303. https://doi.org/10.1103/PhysRevB.82.165303 (2010).
10. Ryazanov, V. V. *et al.* Coupling of two superconductors through a ferromagnet: Evidence for a $\pi$ junction. *Phys. Rev. Lett.* **86**, 2427–2430. https://doi.org/10.1103/PhysRev.96.99 (2001).
11. Robinson, J. W. A., Witt, J. D. S. & Blamire, M. G. Controlled injection of spin-triplet supercurrents into a strong ferromagnet. *Science* **329**, 59–61. https://doi.org/10.1126/science.1189246 (2010).
12. Eschrig, M., Kopu, J., Cuevas, J. C. & Schön, G. Theory of half-metal/superconductor heterostructures. *Phys. Rev. Lett.* **90**, 137003. https://doi.org/10.1103/PhysRevLett.90.137003 (2003).
13. Lutchyn, R. M., Sau, J. D. & Das Sarma, S. Majorana fermions and a topological phase transition in semiconductor–superconductor heterostructures. *Phys. Rev. Lett.* **105**, 077001. https://doi.org/10.1103/PhysRevLett.105.077001 (2010).
14. Tagirov, L. R. Low-field superconducting spin switch based on a superconductor/ferromagnet multilayer. *Phys. Rev. Lett.* **83**, 2058–2061. https://doi.org/10.1103/PhysRevLett.83.2058 (1999).
15. Baek, B., Rippard, W. H., Benz, S. P., Russek, S. E. & Dresselhaus, P. D. Hybrid superconducting-magnetic memory device using competing order parameters. *Nat. Commun.* **5**, 3888. https://doi.org/10.1038/ncomms4888 (2014).
16. Linder, J. & Robinson, J. W. A. Superconducting spintronics. *Nat. Phys.* **11**, 307. https://doi.org/10.1038/nphys3242 (2015).
17. Eschrig, M. Spin-polarized supercurrents for spintronics: a review of current progress. *Rep. Prog. Phys.* **78**, 104501. https://doi.org/10.1088/0034-4885/78/10/104501 (2015).
18. Buzdin, A. I. Proximity effects in superconductor–ferromagnet heterostructures. *Rev. Mod. Phys.* **77**, 935–976. https://doi.org/10.1103/RevModPhys.77.935 (2005).
19. Bergeret, F. S., Volkov, A. F. & Efetov, K. B. Odd triplet superconductivity and related phenomena in superconductor–ferromagnet structures. *Rev. Mod. Phys.* **77**, 1321–1373. https://doi.org/10.1103/RevModPhys.77.1321 (2005).
20. Zhu, Y., Pal, A., Blamire, M. G. & Barber, Z. H. Superconducting exchange coupling between ferromagnets. *Nat. Mater.* **16**, 195. https://doi.org/10.1038/nmat4753 (2017).
21. De Gennes, P. Coupling between ferromagnets through a superconducting layer. *Phys. Lett.* **23**, 10–11. https://doi.org/10.1016/0031-9163(66)90229-0 (1966).
22. Alekseevskij, N., Garifullin, I., Kharakash'yan, E. & Kochelaev, B. Electron paramagnetic resonance for localized magnetic states in the superconducting layer system. *Zh. Eksp. Teor. Fiz.* **72**, 1523–1533 (1977).
23. Kochelaev, B., Tagirov, L. & Khusainov, M. Spatial dispersion of the spin susceptibility of conductivity electrons in superconductors. *Zh. Eksp. Teor. Fiz.* **76**, 578–587 (1979).
24. Khusainov, M. G. Z. Indirect RKKY exchange and magnetic states of ferromagnet-superconductor superlattices. *Eksp. Teor. Fiz.* **109**, 524 (1996).
25. Aristov, D. N., Maleyev, S. V. & Yashenkin, A. G. Rkky interaction in layered superconductors with anisotropic pairing. *Z. Phys. B* **102**, 467–471. https://doi.org/10.1103/PhysRev.96.99 (1997).
26. Di Bernardo, A. *et al.* Nodal superconducting exchange coupling. *Nat. Mater.* **1866**, 1–7. https://doi.org/10.1038/s41563-019-0476-3 (2019).
27. Eschrig, M. Spin-polarized supercurrents for spintronics. *Phys. Today* **64**, 43. https://doi.org/10.1063/1.3541944 (2011).

28. Maekawa, S. & Fukuyama, H. Localization effects in two-dimensional superconductors. *J. Phys. Soc. Jpn.* **51**, 1380. https://doi.org/10.1143/JPSJ.51.1380 (1981).
29. Ma, M. & Lee, P. A. Localized superconductors. *Phys. Rev. B* **32**, 5658–5667. https://doi.org/10.1103/PhysRev.96.99 7 (1985).
30. Dmitrienko, V. E., Ovchinnikova, E. N., Kokubun, J. & Ishida, K. Dzyaloshinskii–Moriya interaction: How to measure its sign in weak ferromagnets?. *JETP Lett.* **92**, 383–387. https://doi.org/10.1134/S0021364010180050 (2010).
31. Liu, D. *et al.* Electronic origin of high-temperature superconductivity in single-layer FeSe superconductor. *Nat. Commun.* **3**, 1–6. https://doi.org/10.1038/ncomms1946 (2012).
32. Okazaki, K. *et al.* Superconductivity in an electron band just above the fermi level: Possible route to BCS–BEC superconductivity. *Sci. Rep.* **4**, 1–6. https://doi.org/10.1038/srep04109 (2014).
33. Melin, R. Superconducting cross-correlations in ferromagnets: implications for thermodynamics and quantum transport. *J. Phys. Condens. Matter* **13**, 6445. https://doi.org/10.1088/0953-8984/13/30/301 (2001).
34. Apinyan, V. & Melin, R. Microscopic theory of non local pair correlations in metallic F/S/F trilayers. *Eur. Phys. J. B* **25**, 373. https://doi.org/10.1140/epjb/e20020042 (2002).
35. H. Jirari, R. M. & Stefanakis, N. Proximity effect in multiterminal hybrid structures. *Eur. Phys. J. B* **31**, 125. https://doi.org/10.1140/epjb/e2003-00016-8 (2003).
36. Buzdin, A. & Daumens, M. Inversion of the proximity effect in hybrid ferromagnet-superconductor-ferromagnet structures. *Europhys. Lett.* **64**, 510. https://doi.org/10.1209/epl/i2003-00252-0 (2003).
37. Melin, R. & Feinberg, D. What is the value of the superconducting gap of a F/S/F trilayer? *Europhys. Lett.* **65**, 96. https://doi.org/10.1209/epl/i2003-10051-1 (2004).
38. Melin, R. Microscopic theory of equilibrium properties of F/S/F trilayers with weak ferromagnets. *Eur. Phys. J. B* **39**, 249. https://doi.org/10.1140/epjb/e2004-00188-7 (2004).
39. Beckmann, D., Weber, H. B. & Löhneysen, H. Evidence for crossed Andreev reflection in superconductor–ferromagnet hybrid structures. *Phys. Rev. Lett.* **93**, 197003. https://doi.org/10.1103/PhysRevLett.93.197003 (2004).
40. Fominov, Y. V., Chtchelkatchev, N. M. & Golubov, A. A. Nonmonotonic critical temperature in superconductor/ferromagnet bilayers. *Phys. Rev. B* **66**, 014507. https://doi.org/10.1103/PhysRev.96.99 8 (2002).
41. Mironov, S. V. & Buzdin, A. Standard, inverse, and triplet spin-valve effects in $F_1/s/f_2$ systems. *Phys. Rev. B* **89**, 144505. https://doi.org/10.1103/PhysRev.96.99 9 (2014).
42. Mironov, S., Mel'nikov, A. & Buzdin, A. Electromagnetic proximity effect in planar superconductor-ferromagnet structures. *Appl. Phys. Lett.* **113**, 022601. https://doi.org/10.1063/1.5037074 (2018).
43. Devizorova, Z., Mironov, S. V., Mel'nikov, A. S. & Buzdin, A. Electromagnetic proximity effect controlled by spin-triplet correlations in superconducting spin-valve structures. *Phys. Rev. B* **99**, 104519. https://doi.org/10.1103/PhysRevLett.61.2472 0 (2019).

### Acknowledgements
This work was supported by the Research Council of Norway through its Centres of Excellence funding scheme grant 262633 QuSpin. We thank L. G. Johnsen for useful discussions. We also thank A. Di Bernardo and O. Millo for very helpful comments.

### Author contributions
J.L. came up with the idea for the project. A.B.G. performed the analytical and numerical calculations with support from V.K.R. All authors contributed to the discussion of the results and the writing of the manuscript.

### Competing interests
The authors declare no competing interests.

### Additional information
**Supplementary information** The online version contains supplementary material available at https://doi.org/10.1038/s41598-021-83620-3.

**Correspondence** and requests for materials should be addressed to J.L.

**Reprints and permissions information** is available at www.nature.com/reprints.

**Publisher's note** Springer Nature remains neutral with regard to jurisdictional claims in published maps and institutional affiliations.